\documentclass[prb,aps,twocolumn]{revtex4}
\usepackage{graphicx,color,latexsym}
\usepackage{dcolumn}
\usepackage{amsmath,amssymb,epsf,bm}
\begin{document}
\title{AC anomalous Hall effect in topological insulator Josephson junctions}
\author{A.~G. Mal'shukov}
\affiliation{Institute of Spectroscopy, Russian Academy of Sciences, Troitsk, Moscow, 108840, Russia}
\affiliation{Moscow Institute of Physics and Technology, Institutsky per.9, Dolgoprudny, 141700 Russia}
\affiliation{National Research University Higher School of Economics, Myasnitskaya str. 20, Moscow, 101000 Russia}
\begin{abstract}
A nonstationary anomalous Hall current is calculated for a voltage biased Josephson junction, which is composed of two  s-wave superconducting contacts deposited on the top of a three dimensional topological insulator (TI). A homogeneous Zeeman field was assumed at the surface of TI. The problem has been considered within the ballistic approximation and on the assumption that tunneling of electrons between contacts and the surface of  TI is weak. In this regime the Josephson current has no features of the $4\pi$-periodic topological effect which is associated with Andreev bound states. It is shown that the Hall current oscillates in time. The phase of these oscillations is shifted by $\pi/2$ with respect to the Josephson current and their amplitude linearly decreases with the electric potential difference between contacts.
It  is also shown that the Hall current cannot be induced by a stationary phase difference of contact's order parameters.

\end{abstract}
\maketitle

\section{Introduction}

Anomalous Hall effect (AHE), as well as a more conventional Hall effect, was first observed by Hall more than a century ago \cite{Hall}. It has been named "anomalous" because in some magnetic materials this effect was detected in the absence of an external magnetic field. In magnetic systems  AHE may be explained by the presence of topologically nontrivial magnetic textures \cite{Matl,Ye,Chun,Taguchi}. Other theories of AHE do not rely on topological magnetic textures, but rather on a strong spin-orbit coupling (SOC) of electrons which, in combination with a homogeneous magnetic order,  can lead to AHE. \cite{AHE rewiev} Recently, great interest to AHE was attracted by a discovery of the quantum anomalous Hall effect \cite{Haldane} which can be observed in one-dimensional  quantized transport of electrons along edges of topological insulators.\cite{QAHE review,Liu2008} This quantized effect has recently been  observed in various systems. \cite{Chang2013,Checkelscy,Kou}

In addition to AHE in normal metals, the anomalous Hall transport in superconducting systems is fundamental to many practical applications. In some cases this effect may be realized in topologically nontrivial materials via the quantized electron transport of Majorana fermions along chiral edge channels. \cite{Qi,Qi Rev.,Wang,Chen1} The  nontopological AHE was also considered. \cite{Sacramento,Yokoyama,Wang1} On the other hand, the latter effect has not been addressed sufficiently, while it can extend considerably functionality of superconducting quantum circuits. Important elements of such circuits are Josephson junctions. A great interest is attracted to junctions where the weak link is represented by a two-dimensional (2D) electron gas on the conducting surface of a  three dimensional (3D) topological insulator.  \cite{Fu,Tkachov,Linder,Morpurgo,Veldhorst,Molenkamp,Kurter,Kayyalha,Mason,Yano,Assouline}
Since such systems are characterized by the strong SOC, it is of fundamental interest to find out wether the  AHE can be observed there together with the Josephson effect.  In contrast to the latter, which manifests itself in a supercurrent between superconducting contacts, the anomalous Hall supercurrent should be directed parallel to the gap separating the contacts. The AHE may be observed only in systems with the broken time inversion symmetry, for example, in the presence of a magnetic order, which adds a mass term into Dirac Hamiltonian of electron states on the surface of TI. Such a magnetic TI can be created by a magnetic impurity doping\cite{Chang2013,Checkelscy,Kou}, in TI magnetic insulator heterostructures, \cite{Otrokov2,Luo,Eremeev,Katmis,He} or in antiferromagnetic TI \cite{Zeugner,Otrokov}.

Our goal is to study the nontopological AHE in Josephson contacts with a magnetic TI taken as a weak link. The  Josephson current in nontopological junctions has a conventional $2\pi$ periodicity, as a function of the phase difference between order parameters of superconducting contacts. In contrast, the topological Josephson effect is characterized by the 4$\pi$ periodicity. \cite{Fu} From the experimental point of view such a  nontopological AHE is of special interest, because it does not require special experimental conditions. In particular, it is not necessary to provide a large proximity induced superconducting gap on the surface of TI, in order to guarantee that Andreev bound states will dominate the electron transport between contacts. We will consider a  junction under the voltage bias, because the analysis below shows that AHE can not be driven by a static phase difference between superconducting terminals. Therefore, the Hall (super)current oscillates in time, as well as the Josephson current of Cooper pairs. The model system is shown in Fig.1. It will be assumed that the superconducting contacts are weakly coupled to the TI surface. Therefore, the superconducting proximity effect, which is induced in the TI by the contacts may be taken into account perturbatively. In this situation the role of Andreev bound states of TI electrons is not important, because the proximity induced  minigap under contacts is small. It is of the order of the tunneling rate $\Gamma$ between a contact and TI. This rate is assumed to be much less than the superconducting gap $\Delta$ in both contacts. Hence, the transport of electrons between superconductors mostly occurs by means of quasiparticle states outside the minigap.

Even within the perturbation theory an analysis of the considered problem poses serious difficulties because it is beyond a conventional semiclassical approach \cite{RevKeldysh}. Moreover, on this reason one can not use the Born approximation for an analysis of impurity scattering effects. Therefore, it will be assumed that the transport of electrons between contacts is ballistic. It requires the sufficiently small distance $L$ between contacts. For example, highly ballistic TI  junctions were reported in Ref. \onlinecite{Kayyalha} with $L\simeq 100 nm$.

The article is organized in the following way. The formalism employed in this work will be presented in Sec. II. In Sec. III two situations will be considered, depending on the position of the chemical potential with respect to the energy gap, which is induced by the Zeeman field. A discussion of results is presented in Sec. IV.

\begin{figure}[tp]
\includegraphics[width=6cm]{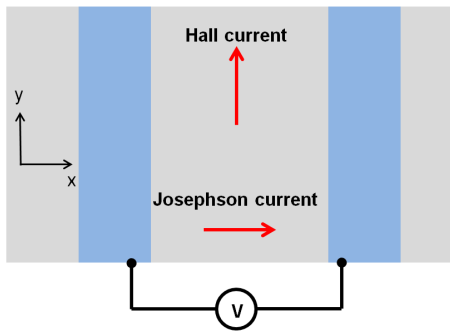}
\caption{(Color online) Two superconducting contacts are placed on the top of a three dimensional magnetic topological insulator. The voltage bias $V$ is applied to the contacts. Under this bias the oscillating in time anomalous Hall current of Cooper pairs is induced perpendicular to the Josephson current.} \label{fig1}
\end{figure}
%%%%%%%%%%%%%%%%%%%%%%%%%%%%%%%%%%%%%%%%%%%%%%%%%%%%
\section{Formalism}

The  unperturbed Hamiltonian of two dimensional (2D) electron gas on the TI surface is given by \cite{Qi Rev.} $H_0=\sum_{\mathbf{k}}\psi^{\dag}_{\mathbf{k}}\mathcal{H}_{0\mathbf{k}}\psi_{\mathbf{k}}$, where $\psi_{\mathbf{k}}$ are the electron field operators,  which are defined in the Nambu basis as $\psi=(\psi_{\uparrow},\psi_{\downarrow},\psi^{\dag}_{\downarrow},-\psi^{\dag}_{\uparrow})$ and the one-particle Hamiltonian
$\mathcal{H}_{0\mathbf{k}}$ is given by
\begin{equation}\label{H0}
\mathcal{H}_{0\mathbf{k}}=\tau_3(v\mathbf{k}\times\bm{\sigma}-\mu)+\tau_0M\sigma_z
\end{equation}
Here, $\mu$ is the  chemical potential, $M$ is the exchange field produced by the magnetic order and $\sigma^j$ denote Pauli matrices ($j=x,y,z$). The Pauli matrices $\tau_i$, $i=0,1,2,3$, operate in the Nambu space, where $\tau_0$ is the unit matrix. We assume a weak tunneling coupling between TI and superconducting contacts. The corresponding tunneling Hamiltonians $H_{L}$ and  $H_{R}$ for the left and right contacts can be written in the form
\begin{equation}\label{Hint}
H_{L(R)}=\sum_{\mathbf{k},\mathbf{k}^{\prime}}(\psi^{\dag}_{\mathbf{k}}t_{L(R)\mathbf{k},\mathbf{k}^{\prime}}\tau_3\psi^{S}_{L(R)\mathbf{k}^{\prime}}+ h.c.)\,,
\end{equation}
where $\psi^{S}_{L(R)\mathbf{k}^{\prime}}$ are electron's field operators in the left and right contacts.  Generally, the tunneling parameters $t_{L(R)\mathbf{k},\mathbf{k}^{\prime}}$ are spin dependent. Let us consider, as an example, time-reversal-symmetric TI belonging to Bi$_2$Se$_3$ family. In these materials in the leading $\mathbf{k}\cdot\mathbf{p}$ expansion the tunneling parameters are diagonal in spin space, but are different for opposite spin projections. This is dictated by the form of Bloch functions which are associated with surface states near the Dirac point. According to Refs. [\onlinecite{Zhang,Qi Rev.}], the degenerate pair of such functions has the form $(\psi_1+i\psi_2)|\uparrow\rangle$ and $(\psi_1-i\psi_2)|\downarrow\rangle$, where the arrows denote the spin projection and real functions $\psi_1$ and $\psi_2$ are composed from $p_z$ atomic orbitals. Within the tight binding approximation the tunneling parameters are determined by the respective overlap integrals $a_1$ and $a_2$ of these functions with adjacent to TI atomic orbitals of a contact material (superconductor, or a spacer). The latter are assumed spin-independent. Therefore, the tunneling parameters in Eq.(\ref{Hint}) are proportional to $a_1 + i a_2$ and $a_1 - i a_2$ for up and down spin projections, respectively. A $\mathbf{k}\cdot\mathbf{p}$ expansion near the $\Gamma$-point may result in small spin-dependent corrections, which will be ignored below. One should take into account that, since the contact size in the $x$-direction is finite, the in-plane component of the wave vector of a tunneling particle is not conserved. Therefore, $t_{L(R)\mathbf{k},\mathbf{k}^{\prime}} \propto \delta_{\mathbf{k}-\mathbf{k}^{\prime},\mathbf{q}}$,  where $\mathbf{q}$ is the Fourier wave vector of a function which describes the contact shape.

The Hall current in the junction is directed parallel to the $y$-axis. The corresponding one-particle current operator is given by $v\sigma_x$. Therefore, the Hall current density $J_H^y$ may be expressed in terms of the Keldysh Green function as
\begin{equation}\label{JH}
J_{H}^y(\mathbf{r},t)=-\frac{ive}{4}\mathrm{Tr}[\sigma^x
G^K(t,\mathbf{r};t,\mathbf{r})]\,.
\end{equation}
The Josephson current and anomalous Hall current are given by the fourth order in the expansion of $G^K$ with respect to the tunneling parameter. Such a perturbative approach was previously employed for calculation of  the Josephson current \cite{Aslamazov} and the spin-Hall current in voltage biased Josephson junctions \cite{Malshukov}. Each superconducting contact gives rise to the self-energy $\Sigma_{L(R)}$, which may be written in the form
 \begin{equation}\label{Sigma}
\Sigma_{L(R)\mathbf{k},\mathbf{k+q}}(t,t^{\prime})=\sum_{\mathbf{k}^{\prime}}t_{L(R)\mathbf{k},\mathbf{k}^{\prime}}
t_{L(R)\mathbf{k}^{\prime},\mathbf{k}+\mathbf{q}}
G^{S}_{\mathbf{k}^{\prime}}(t,t^{\prime})\,.
\end{equation}
Because of the electric potentials  $V_{L/R}/e=\pm V/2e$ on contacts, $\Sigma_{L(R)\mathbf{k},\mathbf{k+q}}(t,t^{\prime})$ takes the form
\begin{equation}\label{Sigmatt}
\Sigma_{L(R)\mathbf{k},\mathbf{k+q}}(t,t^{\prime})= e^{i\tau_3V_{L(R)}t} \Sigma_{L(R)\mathbf{k},\mathbf{k+q}}(t-t^{\prime})e^{-i\tau_3V_{L(R)}t^{\prime}}\,.
\end{equation}

Although $t_{L(R)\mathbf{k},\mathbf{k}^{\prime}}$ depends on spin, the self-energy in  Eq.(\ref{Sigma}) is a spin independent function. It is guaranteed by a spin-singlet structure of the superconductor Green's function $G^S$ in Eq.(\ref{Sigma}) and by a form of the spin dependence of tunneling parameters. Indeed, since the spin dependent part of $t_{\mathbf{k},\mathbf{k}^{\prime}}$ has the form $a_1 \pm i a_2$ with real $a_1$ and $a_2$, it is easy to see that for both spin orientations the self energy will be proportional to $a_1^2 +a_2^2$ and, hence, the self energy is spin independent.

 As was discussed above, the only important wave-vector dependence of $t_{L(R)\mathbf{k},\mathbf{k}^{\prime}}$ is associated with the finite size of contacts in the $x$-direction. The shape of the contacts may be taken into account by multiplying the self energies $\Sigma_{L(R)}$ by the functions $s_{L}(x)=\theta(x-x_{L2})\theta(x_{L1}-x)$ and $s_{R}(x)=\theta(x-x_{R1}\theta(x_{R2}-x)$ where the distance between the contacts is $L=x_{R1}-x_{L1}$. The Fourier components of these functions will be denoted as $s_{L(R)}(\mathbf{q})$. By integrating  $G^S$ in Eq.(\ref{Sigma}) over $\mathbf{k}^{\prime}$ one may obtain a simple expression for  temporal Fourier components of the retarded (r) and advanced (a) self energies $\Sigma^{r/a}_{L(R)}(t-t^{\prime})$ in the form
\begin{equation}\label{SigmaW}
\Sigma^{r/a}_{L(R)\mathbf{k},\mathbf{k+q}}(\omega)=\Gamma s_{L(R)}(\mathbf{q})\frac{\tau_1\Delta}{\sqrt{(\omega
\pm i\delta)^2-\Delta^2}}\,,
\end{equation}
where $\Delta$ is the superconducting order parameter and $\Gamma$ can be expressed  through the resistance $R_b$ of the SN interface, as  $\Gamma=1/4e^2 N_F R_b$, with $N_F$ denoting the state density at the Fermi energy. Only nondiagonal matrix elements of $\Sigma$ in the Nambu space are taken into account, because only these terms contribute to the Josephson current in the tunneling regime. In the following it will be convenient to write Nambu matrix elements of $\Sigma^{r/a}_{L(R)\mathbf{k},\mathbf{k+q}}(\omega)$ in Eq.(\ref{SigmaW}) as
\begin{equation}\label{SigmaW2}
\left(\Sigma^{r/a}_{L(R)\mathbf{k},\mathbf{k+q}}(\omega)\right)_{12}=\left(\Sigma^{r/a}_{L(R)\mathbf{k},\mathbf{k+q}}(\omega)\right)_{21}=
\Sigma^{r/a}(\omega)s_{m\mathbf{q}}\,,
\end{equation}
where the subscript $m$ takes the values 1 and -1 for $\Sigma_R$ and $\Sigma_L$, respectively, and $\Sigma^{r/a}(\omega)=\Gamma\Delta/\sqrt{(\omega
\pm i\delta)^2-\Delta^2}$.

We are interested in the total electric current in the y-direction. Therefore, the current density in Eq.(\ref{JH}) should be integrated over $x$. In the second order with respect to $\Gamma$ Fourier components of the total current can be expressed from Eq.(\ref{JH}) in the form
\begin{eqnarray}\label{JHtot}
&&J_{H}^y(\pm 2\Omega)=ie\frac{v}{4}
\int \frac{d\omega}{2\pi} dxdx^{\prime}\sum_{\mathbf{k},\mathbf{q},m=\pm\tau}s_{m}(x)s_{-m}(x^{\prime})\times
\nonumber \\ &&\mathrm{Tr}\left[\sigma_x G_{\tau,\mathbf{k}+\mathbf{q}}(\omega+m\tau\Omega)\Sigma\left(\omega+\frac{m\tau\Omega}{2}\right)
G_{-\tau,\mathbf{k}}(\omega)\right. \times \nonumber \\
&& \left.\Sigma\left(\omega-\frac{m\tau\Omega}{2}\right) G_{\tau,\mathbf{k}+\mathbf{q}}(\omega-m\tau\Omega)\right]^K e^{iq(x^{\prime}-x)},
\end{eqnarray}
where the superscript $K$ denotes the Keldysh component of the matrix product in square brackets, $\Omega=V$, $q\equiv q_x$, and the trace is taken over spin variables. The unperturbed Green functions $G_{\tau,\mathbf{k}}(\omega)$ of TI electrons in Eq.(\ref{JHtot}) may be obtained from  Eq.(\ref{H0}), where $\tau_3 \rightarrow \tau=\pm 1$. Therefore, the corresponding  retarded, advanced and Keldysh functions are given by
\begin{eqnarray}\label{G}
G^{r(a)}_{\tau,\mathbf{k}}(\omega)&=&(\omega-\tau v\mathbf{k}\times\bm{\sigma}+\tau\mu-M\sigma_z \pm i\delta)^{-1}\,,\nonumber \\
G^{K}_{\tau,\mathbf{k}}(\omega)&=&(G^{r}_{\tau,\mathbf{k}}(\omega)-G^{a}_{\tau,\mathbf{k}}(\omega))\tanh\frac{\omega}{2k_B T}\,,
\end{eqnarray}
where $T$ is the temperature. Since the system is uniform in the $y$-direction, one may set $k_y=0$. Therefore, the Green functions in Eq.(\ref{G}) can be written as
\begin{equation}\label{G2}
G^{r(a)}_{\tau,\mathbf{k}}(\omega)=\frac{\omega+\tau vk_x\sigma_y+\tau\mu+M\sigma_z}{(\omega+\tau\mu\pm i\delta)^2-E_{\mathbf{k}}^2} \,,
\end{equation}
where $E_{\mathbf{k}}=\sqrt{v^2k_{x}^2+M^2}$ and $\delta\rightarrow 0$.

The sum over $\mathbf{k}$ and $\mathbf{q}$ in Eq.(\ref{JHtot}) involves a product of three Green functions. By taking the Keldysh component of the matrix in Eq.(\ref{JHtot}) one obtains various combinations of the retarded and advanced Green functions. These combinations are summed up over $\mathbf{k}$, $\mathbf{q}$ and  spin variables.  As a result, we obtain a set of the functions $b^{\text{abc}}_{m,\tau}(x-x^{\prime})$, which are given by
\begin{eqnarray}\label{b}
&&b^{\text{abc}}_{m,\tau}=
\sum_{\mathbf{k},\mathbf{q}}\mathrm{Tr}\left[\sigma_x G_{\tau,\mathbf{k}+\mathbf{q}}^{\text{a}}(\omega+m\tau\Omega)\times \right.
\nonumber \\ &&
\left. G^{\text{b}}_{-\tau,\mathbf{k}}(\omega) G^{\text{c}}_{\tau,\mathbf{k}+\mathbf{q}}(\omega-m\tau\Omega)\right]e^{iq(x^{\prime}-x)} \,.
\end{eqnarray}
Each of the symbols $\text{a,b,c}$ takes the values $r$, or $a$.  By taking the trace in Eq.(\ref{b}), the latter can be transformed to
\begin{eqnarray}\label{b2}
&&b^{\text{abc}}_{m,\tau}=-4im\Omega vM
\sum_{\mathbf{k},\mathbf{q}}(2k_x+q)D_{\tau,\mathbf{k}+\mathbf{q}}^{\text{a}}(\omega+m\tau\Omega)\times
\nonumber \\ &&
D^{\text{b}}_{-\tau,\mathbf{k}}(\omega) D^{\text{c}}_{\tau,\mathbf{k}+\mathbf{q}}(\omega-m\tau\Omega)e^{iq(x^{\prime}-x)} \,,
\end{eqnarray}
where the functions $D^r$ and $D^a$ are given by $D^{\text{r/a}}_{\tau,\mathbf{k}}(\omega)=\left((\omega+\tau\mu\pm i\delta)^2-E_{\mathbf{k}}^2\right)^{-1}$. It turned out that the functions $b^{\text{abc}}_{m,\tau}$ are proportional to $\Omega$. This means that within the considered model the Hall current cannot be induced by a "phase" bias, provided by a static phase difference of order parameters in contacts. It follows from Eq.(\ref{b2}) that the functions $b^{\text{abc}}$ satisfy the equations
\begin{eqnarray}\label{bmtau}
&&b^{\text{\text{abc}}}_{m,\tau}(\omega,x)=-b^{\text{\text{abc}}}_{m,\tau}(\omega,-x)\,,\nonumber \\
&&b^{\text{\text{abc}}}_{1,-1}(\omega,x)=b^{*\text{\text{abc}}}_{1,1}(-\omega,x)\,.
\end{eqnarray}

By calculating the Keldysh projection of the matrix in Eq. (\ref{JHtot}) and taking into account Eq.(\ref{bmtau}) the Hall current $J_{H}^y(2\Omega)$ may be written in terms of the functions $b^{\text{abc}}$, as
\begin{equation}\label{JHtot2}
J_{H}^y(2\Omega)=ie\frac{v}{4}
\int \frac{d\omega}{2\pi} dxdx^{\prime}s_{R}(x)s_{L}(x^{\prime})(S_{1}+S_{2}+S_{3})\,,
\end{equation}
where
\begin{eqnarray}\label{S2}
S_{1}=b^{rrr}\Sigma_+^r\Sigma_-^r\tanh\frac{\omega-\Omega}{2k_BT}&-&b^{aaa}\Sigma_+^a\Sigma_-^a
\tanh\frac{\omega+\Omega}{2k_BT}\,,\nonumber \\
S_{2}=b^{rra}\Sigma_+^r\Sigma_-^r\left (\tanh\frac{\omega}{2k_BT}\right.&-&\left.\tanh\frac{\omega-\Omega}{2k_BT}\right)+ \nonumber \\
b^{raa}\Sigma_+^a\Sigma_-^a\left (\tanh\frac{\omega+\Omega}{2k_BT}\right.&-&\left.\tanh\frac{\omega}{2k_BT}\right)\,,
\end{eqnarray}
and
\begin{eqnarray}\label{S3}
&&S_{3}=b^{rra}\Sigma_+^r(\Sigma_-^r-\Sigma_-^a)\left(\tanh\frac{2\omega-\Omega}{4k_BT}-\tanh\frac{\omega}{2k_BT}\right) +\nonumber \\
&&b^{raa}(\Sigma_+^r-\Sigma_+^a)\Sigma_-^a\left(\tanh\frac{2\omega+\Omega}{4k_BT}-\tanh\frac{\omega}{2k_BT}\right)\,,
\end{eqnarray}
where $\Sigma^{r/a}_{\pm}=\Sigma^{r/a}\left(\omega\pm\frac{\Omega}{2}\right)$ and $b^{\text{abc}}=b^{\text{abc}}_{1,1}(x-x^{\prime})+b^{\text{abc}}_{-1,-1}(x^{\prime}-x)$.

In its turn, the time dependence of the Hall current is given by
\begin{equation}\label{Jtime}
J_{H}^y(t)=J_{H}^y(2\Omega)e^{2i\Omega t}+J_{H}^{*y}(2\Omega)e^{-2i\Omega t}
\end{equation}

\section{limiting cases}

\subsection{The chemical potential outside the mass gap}

The exchange interaction $M$ gives rise to a gap in the spectrum of electron states, as can be seen from the poles of the Green function in  Eq.(\ref{G2}). In doped TI the Fermi level can be outside the gap. Let us consider the case when $\mu>M>0$. It will also be assumed that $V\ll \Delta \ll \mu$ and $|q|\ll k_F$, where $vk_F=\sqrt{\mu^2-M^2}$, with the Fermi velocity $v_F$ given by $v_F=v\sqrt{\mu^2-M^2}/\mu$. At these assumptions the functions $b^{\text{abc}}$ can be calculated analytically from Eqs.(\ref{b}) and (\ref{b2}), by linearizing the denominators of Green functions Eq.(\ref{G2}) near the Fermi energy. As a result, we obtain for $m=\tau=1$ and $x-x^{\prime}>0$
\begin{eqnarray}\label{bfin}
&&b^{rrr}_{1,1}(\omega)=\gamma \int_{-\pi}^{\pi} \frac{d\phi}{2\pi}\exp\frac{2i\omega(x-x^{\prime})}{v_x}\sin\frac{\Omega(x-x^{\prime})}{v_x}\,,\nonumber \\
&&b^{aaa}_{1,1}(\omega)=b^{rrr*}_{1,1}(\omega)\,,\nonumber \\
&&b^{rra}_{1,1}(\omega)=-i\frac{\gamma}{2} \int_{-\pi}^{\pi} \frac{d\phi}{2\pi}\exp\frac{2i\omega(x-x^{\prime})}{v_x}\exp\frac{i\Omega(x-x^{\prime})}{v_x}\,,\nonumber \\
&&b^{raa}_{1,1}(\omega)=b^{rra}_{1,1}(-\omega)\,, b^{\text{abc}}_{1,1}(x-x^{\prime})=b^{\text{abc}}_{-1,-1}(x^{\prime}-x) \,,
\end{eqnarray}
where $\gamma=2 M/\mu v^3$ and $v_x=v_F\cos\phi >0$.
\begin{figure}[tp]
\includegraphics[width=6cm]{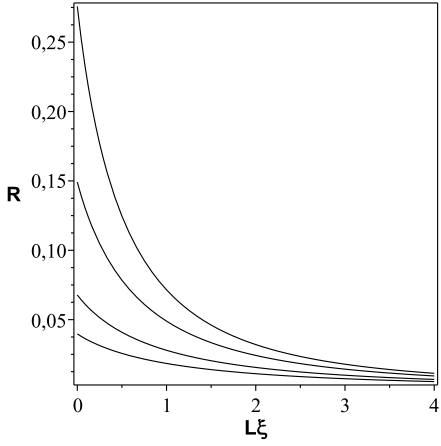}
\caption{The anomalous Hall current as a function of the distance between contacts (see Eq. (\ref{JH12})). Curves from the top to the bottom: $w\xi$=0.5, $w\xi$=1, $w\xi$=2 and $w\xi$=3, where $\xi=\Delta/v$} \label{fig2}
\end{figure}

According to Eq. (\ref{SigmaW}), at $\Omega\ll \Delta$ the function $\Sigma_{\pm}^r-\Sigma_{\pm}^a$ in Eq. (\ref{S3}) is finite only at $\omega \gtrsim \Delta$. In this range, however, the temperature dependent  factors $\left[\tanh(\omega/2k_BT\pm\Omega/2k_BT)-\tanh(\omega/2k_BT)\right]$ are exponentially small, as $\exp(-2\Delta/k_BT)$ at $k_BT \ll \Delta$. Therefore, one may ignore $S_3$ in Eq. (\ref{JHtot2}). An important parameter range is determined by the distance $L$ between contacts and by their width $w$.  In the case of  $w \sim L$ the characteristic flight time  $T_f =L/v$. For a typical TI (for example, Bi$_2$S$_3$)  with $v=5\cdot 10^5$m/s and  $L=100$nm one obtains $1/T_f=3$meV. Therefore, for such a ballistic junction both $1/T_f \gg \Omega$ and $\Delta  \gg \Omega$. In this case a contribution to Eq. (\ref{JHtot2}) given by $S_2$ can be easily calculated, because, according to Eq. (\ref{S2}), only small $\omega \sim \Omega$ contribute to $S_2$. Hence, one may set $\omega=0$ in $b^{rra}$ and $b^{raa}$ in Eq. (\ref{bfin}), as well as in $\Sigma^{r/a}$ given by Eqs. (\ref{SigmaW}) and (\ref{SigmaW2}). In this case $\Sigma_{\pm}^{r/a}=i\Gamma$ in Eq.(\ref{S2}). As a result, the integration over $x,x^{\prime}, \omega$ and $\phi$ in Eqs. (\ref{JHtot2}) and (\ref{bfin}) gives for the current $J_{H2}^y(2\Omega)$, which is associated with the second term in  Eq. (\ref{JHtot2}), the following expression
\begin{equation}\label{JH2}
J_{H2}^y(2\Omega)=\frac{e}{4\pi}\Omega\Gamma^2 w^2 \frac{M}{\mu v^2}
\end{equation}
The total current, which includes the first two terms in  Eq. (\ref{JHtot2}) and ignores the small third one, can be written in the form
\begin{equation}\label{JH12}
J_{H}^y(2\Omega)=J_{H2}^y(2\Omega)R(L,w)\,,
\end{equation}
where the function $R$ is plotted as a function of $L$ in Fig. 2, at various $w$.

 \subsection{The chemical potential inside the mass gap}

 In this section we will assume $\mu=0$. In this case the magnetic gap $M$ prevents penetration of Cooper pairs into TI. Therefore, the distance between superconducting contacts must be small enough, so that $MT_f \lesssim 1$. In the same way as in Eq. (\ref{JHtot2}), the anomalous Hall current  can be expressed in terms of the functions $b^{\text{abc}}$, that are given by Eq. (\ref{b2}). In contrast to Sec.IIIA, however, the factors $b^{rra}$ and $b^{raa}$ are proportional to $\Omega$ at $\Omega \ll M$. Therefore, by taking into account the temperature dependent statistical factor in Eq. (\ref{S2}), which is $\sim \Omega$, we arrive to $S_2 \sim \Omega^2$. Hence, the leading contribution to $J_H$ is given by $S_1$. Since $b^{rrr}\Sigma_+^r\Sigma_-^r$ and $b^{aaa}\Sigma_+^a\Sigma_-^a$ are analytical functions of $\omega$ in the upper and lower complex semiplanes, respectively, it is convenient to transform the integration over $\omega$ in  Eq. (\ref{JHtot2}) into the sum over Matsubara frequencies $\Omega_n=\pi(2n+1)$. As a result, in the leading approximation with respect to $\Omega$ Eq. (\ref{JHtot2}) gives
 \begin{eqnarray}\label{JHtot3}
&&J_{H}^y(2\Omega)=ie2\Omega v^2M T\Gamma^2\int  dxdx^{\prime}s_{R}(x)s_{L}(x^{\prime}) \nonumber \\ &&\sum_{\omega_n,\mathbf{k},q}\frac{2k_x-q}{(\omega_n^2+E_{\mathbf{k}}^2)^2}\frac{\exp[ iq(x^{\prime}-x)]}{\omega_n^2+
E_{\mathbf{k-q}}^2}\frac{\Delta^2}{\omega_n^2+\Delta^2}
 \end{eqnarray}
where $\mathbf{q}=q\mathbf{n}_x$. Analytical expressions for the Hall current may be obtained, by assuming that the distance between contacts $L\ll v/M$. Since at $v/M <L$ the current decreases fast, this limiting case gives the upper bound on the current. In order to analyze main qualitative trends, it is sufficient to consider the cases of the wide ($w\gg v/M$) and narrow ($w\ll v/M$) contacts. At small temperatures $T \ll \Delta$ Eq. (\ref{JHtot3}) gives
\begin{equation}\label{JH3}
J_{H}^y(2\Omega)=e\frac{\Omega}{2\pi}\frac{\Delta}{24}\frac{\Gamma^2}{M^2} \frac{\Delta+2M}{(\Delta+M)^2}
\end{equation}
for wide contacts, and
\begin{equation}\label{JH4}
J_{H}^y(2\Omega)=e\frac{\Omega}{2\pi}\frac{w^2}{4}\frac{\Gamma^2}{v^2} \frac{\Delta}{\Delta+M}
\end{equation}
for narrow contacts.
\section{Discussion}

The anomalous Hall current has been calculated for a ballistic Josephson junction in two regimes, depending on a position of the chemical potential with respect to the mass gap, which in turn is induced by an exchange field. In both cases, according to Eqs. (\ref{JH2}), (\ref{JH3}) and (\ref{JH4}), the  Hall current is proportional to $V\cos 2Vt$. Hence, its oscillation amplitude vanishes at $V\rightarrow 0$. This result also signals that the Hall effect can not be observed in a stationary, where the Josephson current  is induced by the phase difference between  superconducting contacts. This effect has no relevance to the topological 4$\pi$ Josephson tunneling \cite{Fu}. The latter requires a good contact of superconductors with the TI surface, so that a sufficiently large proximity induced energy gap might be formed under contacts. Such gap can support  the bound Andreev states which are involved into the $4\pi$ Josephson current. In the considered here case, however, such a proximity gap is equal to the small tunneling rate $\Gamma$ and can be ignored.

Above, the Josephson current was calculated within the ballistic approximation. It is important to understand a possible influence of  disorder on this current. At the first sight it  seems that this influence is weak when the mean free path $l$ of electrons in TI is much larger than  the distance between contacts, as well as their size in the $x$-direction. The situation, however, is more complicated in the case when $\mu>M$. It becomes evident from an analysis of the Hall current distribution in the $x$-direction. As shown in Appendix A, the current which is associated with $S_2$ in Eq. (\ref{JHtot2}) extends far outside the contacts over the distance $\sim v/\Omega$. At small  $\Omega$ the latter can exceed $l$ and the ballistic approximation becomes invalid. It is reasonable to expect that at such small frequencies the Hall current is able to penetrate only over the distance which is less than $l$.  As a result,  at $\sim v/\Omega \gg l$ the Hall current $J_{H2}^y$ should be proportional to $\sim \Omega^2$, rather than $\Omega$ in Eq. (\ref{JH2}). At the same time, as shown in Appendix A, the Hall current which is associated with $S_1$ is distributed only in the region between and under contacts. Therefore, the effect of a disorder is not so destructive on this current, as long as the contacts are close to each other and are not too wide.

In the case of $\mu< M$ the mass gap restricts the distance over which the current propagates outside the contacts. Hence, the scattering effects are not  important, as long as $M \gg v/l$.

In the considered here model the contacts are infinitely extended in the $y$-direction. A restriction of their size in this direction would lead to charge accumulation and to an electric potential  buildup near contact ends. Therefore, the presence of low ohmic normal contacts is assumed at $y=\pm \infty$. Also, one may consider TI wires, or ribbons, which are coated with superconducting films. In this case the Hall current will circulate around the wire.

\emph{Acknowledgements} - The work was partly supported by the Russian Academy of Sciences program "Actual
 problems of low-temperature physics."
%%%%%%%%%%%%%%%%%%%%%%%%%%%%%%%%%%%%%%%%%%%%%%%%%%%%
%%%%%%%%%%%%%%%%%%%%%%%%%%%%%%%%%%%%%%%%%%%%%%%%%%%%%%%%%

%%%%%%%%%%%%%%%%%%%%%%%%%%%%%%%%%%%%%%%%%%%%%%
\appendix

\section{Spatial distribution of the Hall current density}

The above analysis has been focused on a calculation of the total anomalous Hall current, which is given by an integral of the current density over the $x$ coordinate. On the other hand, the $x$ dependence of this density allows to better understand  a physics of the considered Hall effect.  The current density $J_{H}^y(\pm 2\Omega,x)$ has a more complicated structure in comparison with Eq. (\ref{JHtot}) and is given by
\begin{widetext}
\begin{eqnarray}\label{JHtotdensity}
&&J_{H}^y(\pm 2\Omega,x)=\frac{iev}{4}
\int \frac{d\omega}{2\pi} dx^{\prime}dx^{\prime\prime}\sum_{m=\pm\tau}s_{m}(x^{\prime})s_{-m}(x^{\prime\prime})\sum_{\mathbf{k},\mathbf{q}_1,\mathbf{q}_2}e^{iq_1(x-x^{\prime})}e^{iq_2(x-x^{\prime\prime})}\times
\nonumber \\ &&\mathrm{Tr}\left[\sigma_x G_{\tau,\mathbf{k}+\mathbf{q}_1}(\omega+m\tau\Omega)\Sigma\left(\omega+\frac{m\tau\Omega}{2}\right)
G_{-\tau,\mathbf{k}}(\omega)\Sigma\left(\omega-\frac{m\tau\Omega}{2}\right) G_{\tau,\mathbf{k}-\mathbf{q}_2}(\omega-m\tau\Omega)\right]^K\,.
\end{eqnarray}
\end{widetext}
Instead of the functions $b^{\text{abc}}_{m,\tau}$, which are defined by Eq. (\ref{b}), one can introduce the $x$-dependent functions $b^{\text{abc}}_{m,\tau}(x)$. In the range of parameters considered in Sec. IIIA they have a form
\begin{eqnarray}\label{b2x}
&&b^{\text{abc}}_{m,\tau}(x)=-8im\Omega vM
\sum_{\mathbf{k},\mathbf{q}_1,\mathbf{q}_2}k_xe^{iq_1(x-x^{\prime})}e^{iq_2(x-x^{\prime\prime})}\times
\nonumber \\
&&D_{\tau,\mathbf{k}+\mathbf{q}_1}^{\text{a}}(\omega+m\tau\Omega)D^{\text{b}}_{-\tau,\mathbf{k}}(\omega)
D^{\text{c}}_{\tau,\mathbf{k}-\mathbf{q}_2}(\omega-m\tau\Omega)
 \,.
\end{eqnarray}
From this equation one obtains
\begin{eqnarray}\label{bfinx}
b^{rrr}_{1,1}(\omega)&=&\gamma \theta(x^{\prime}-x) \theta(x-x^{\prime\prime})\Omega\Phi(x)\,, \nonumber \\
b^{rra}_{1,1}(\omega)&=&-\gamma \theta(x^{\prime}-x^{\prime\prime}) \theta(x^{\prime\prime}-x)\Omega\Phi(x) \,,\nonumber \\
b^{raa}_{1,1}(\omega)&=&-\gamma\theta(x^{\prime}-x^{\prime\prime}) \theta(x-x^{\prime})\Omega\Phi^*(x) \,,\nonumber \\
b^{aaa}_{1,1}(\omega)&=&b^{rrr*}_{1,1}(\omega)\,,
\end{eqnarray}
where $\theta(x)$ is the Heaviside step function and
\begin{equation}\label{Fhi}
\Phi(x)=\int_{-\pi}^{\pi} \frac{d\phi}{2\pi}e^{2i\widetilde{\omega}(x^{\prime}-x^{\prime\prime})}e^{-2i\widetilde{\Omega} x}e^{i\widetilde{\Omega}(x^{\prime}+x^{\prime\prime})}\,
\end{equation}
In these equations  $\widetilde{\Omega}=\Omega/v_x$ and $\widetilde{\omega}=\omega/v_x$. Other functions, such as  $b^{\text{abc}}_{1,-1}$, depend on $x$ in a similar way.

It is easy to see that the integration of Eq. (\ref{bfinx}) over $x$ from $-\infty$ to $+\infty$ results in Eq. (\ref{bfin}). However, the convergence of integrals for the functions $b^{rrr}$ and $b^{aaa}$ is much better in comparison with the convergence of $b^{rra}$ and $b^{raa}$. In the former case $x$ is confined due to theta-functions between $x^{\prime}$ and $x^{\prime\prime}$, which are coordinates belonging to the contacts, while in the latter case $x$ is free to vary either to $+\infty$ (for $b^{raa}$) or to $-\infty$ (for $b^{rra}$). Therefore, in this case the corresponding integrals are converging only at finite $\Omega$. This is the consequence of the ballistic approximation used in this work. As a result, there are physical restrictions on $\Omega$. Namely, this frequency must be large enough with respect to various competing parameters which may restrict the distance of the ballistic propagation of electrons. For example, it must be large in comparison with the elastic scattering rate. Since, according to  Eq. (\ref{S2}), $b^{rra}$ and $b^{raa}$ contribute to $J_{H2}$ in Eq. (\ref{JH2}), this current, must depend strongly on the impurity scattering, as it is discussed in Sec. IV.
%%%%%%%%%%%%%%%%%%%%%%%%%%%%%%%%%%%%%%%%%%%%%%%%
\end{document}